\documentclass[article]{aa}
\providecommand{\linenumbers}{}
\providecommand{\nolinenumbers}{}
\providecommand{\switchlinenumbers}{}
\usepackage[varg]{txfonts} 
\usepackage{graphicx}
\usepackage{physics}
\usepackage{mathrsfs}

\usepackage{natbib}
\bibpunct{(}{)}{;}{a}{}{,}

\usepackage{xcolor} 
\usepackage{hyperref}
\hypersetup{
  colorlinks=true,
  linkcolor=red!50!black,
  citecolor=blue!50!black,
  urlcolor=blue!80!black
}

\addto\extrasenglish{%
}

\usepackage{soul}
\usepackage[normalem]{ulem} 

\newif\ifinternal
\internalfalse     

\definecolor{deeppink}{rgb}{1.0, 0.08, 0.58}

\newif\ifrev
\revfalse          

\newcommand\bb[1]{\boldsymbol{#1}}

\begin{document}
\nolinenumbers
\renewcommand\thelinenumber{}

\title{Plasma Mixing Driven by the \\Collisionless Kelvin--Helmholtz Instability}
\subtitle{Insights from fully kinetic simulation and density-based diagnostics}


\author{
Silvia Ferro\inst{1}\fnmsep\thanks{silvia.ferro@kuleuven.be}
\and
Fabio Bacchini\inst{1,2}
\and
Giuseppe Arrò\inst{3}
\and
Francesco Pucci\inst{4}
\and
Pierre Henri\inst{5,6}
}

\institute{
Centre for mathematical Plasma Astrophysics, Department of Mathematics, KU Leuven, 
Celestijnenlaan 200B, B-3001 Leuven, Belgium
\and
Royal Belgian Institute for Space Aeronomy, Solar-Terrestrial Centre of Excellence, Ringlaan 3, 1180 Uccle, Belgium
\and 
Department of Physics, University of Wisconsin-Madison, Madison, WI 53706, USA
\and
Institute for Plasma Science and Technology, National Research Council, CNR-ISTP, Bari, Italy
\and
Laboratoire Lagrange, Observatoire Côte d'Azur, Université Côte d'Azur, CNRS, Nice, France
\and
LPC2E, CNRS, Université d'Orléans, CNES, Orléans, France
}

\abstract
   {Simulations and observations of the low-latitude magnetosphere–magnetosheath boundary layer indicate that the Kelvin--Helmholtz instability (KHI) drives vortex structures that enhance plasma mixing and magnetic reconnection, influencing transport and particle acceleration.}
   {We investigate the spatial localization, species dependence, and physical mechanisms of plasma mixing driven by the nonlinear evolution of the KHI.}
   {We perform high-resolution two-dimensional Particle-In-Cell simulations using a finite-Larmor-radius shear-flow initial configuration. Plasma mixing is quantified using particle labeling, a complementary density-based mixing tracer, and diagnostics of magnetic reconnection.}
   {Mixing across the shear layer is present but localized, occurring mainly in narrow interface regions and plasma structures. Ions mix more effectively than electrons, which remain largely frozen to field lines. Enhanced mixing spatially and temporally correlates with localized magnetic reconnection within and between KH vortices.}
   {Cross-boundary transport driven by the kinetic KHI remains intrinsically localized and is mediated by vortex advection and magnetic reconnection. Electron mixing is strongly constrained, indicating that kinetic-scale transport across collisionless shear layers remains limited.}


   \titlerunning{Plasma Mixing in Kinetic KHI 
   }
   \authorrunning{Ferro et al.}
    
   \maketitle

\section{Introduction} 
The Kelvin--Helmholtz instability (KHI) can develop when the velocity shear exceeds the Alfv\'en speed associated with the magnetic-field component parallel to the shear
\citep{chandrasekhar1961}. 
At Earth’s low-latitude magnetopause, this condition is frequently met during intervals of strongly northward or southward interplanetary magnetic field (IMF), and signatures of surface waves and rolled-up vortices have been widely observed \citep{hasegawa2004,eriksson2016,stawarz2016,settino2024}. KH activity has also been reported at other planetary magnetospheres, including Mercury (\citealt{sundberg2012,aizawa2020b}), Mars (\citealt{ruhunusiri2016,koh2025}), Saturn (\citealt{ma2015,dialynas2018}), and Jupiter (\citealt{ranquist2019,montgomery2023}), raising the question of how magnetospheric size influences the multi-scale development of the instability.

The KHI is also ubiquitous in heliospheric and astrophysical shear flows. 
In the heliosheath and heliotail, global magnetohydrodynamic (MHD) simulations have shown that velocity shear can drive KH and related instabilities that contribute to large-scale turbulence and plasma transport (e.g., \citealt{kieokaew2021,ma2025}). 
Similar shear-driven processes are expected to operate in magnetized astrophysical jets and other collisionless boundary layers, where the nonlinear evolution of KH vortices may influence magnetic-field restructuring and energy dissipation (\citealt{chow2023b,chow2023a}). 
These large-scale studies highlight the global impact of shear-driven instabilities, but do not resolve the kinetic-scale processes that regulate particle interpenetration within vortices.

After saturation, KH vortices generate strong gradients in velocity, density, and magnetic field down to ion and electron scales, triggering secondary instabilities that broaden the mixing layer and enhance plasma transport across the solar wind--magnetosphere boundary (\citealt{nykyri2001,hasegawa2004,nakamura2017}). Kinetic studies and in-situ observations show that this transport is mediated by processes such as reconnection, wave--particle interactions, and vortex-induced reconnection (VIR; e.g., \citealt{nakamura2011}), enabling solar-wind plasma entry even under northward IMF conditions (\citealt{nykyri2006b,stawarz2016,nakamura2017a,nakamura2020}).
While MHD simulations capture the global KH dynamics and reconnection-driven transport (\citealt{faganello2012a,ma2017,ferro2024}), they do not resolve the kinetic mechanisms governing particle diffusion and heating.
Despite these advances, quantitatively characterizing plasma mixing in fully kinetic KHI remains challenging, particularly in distinguishing genuine plasma interpenetration from large-scale advection (\citealt{nakamura2013,karimabadi2013}).

Several studies have addressed plasma transport and mixing in KH dynamics using both simulations and observations. 
In numerical models, distinguishing plasma populations originating on opposite sides of the shear layer has been achieved through population tagging or passive tracers, to quantify boundary-layer broadening and reconnection-driven transport (e.g., \citealt{nakamura2014,ma2017}). In spacecraft observations, where particle origin cannot be directly tracked, mixing parameters have been introduced based on energy-separated magnetospheric and magnetosheath populations (\citealt{settino2022,settino2024,radhakrishnan2024}). These observational diagnostics provide valuable classification tools for KH-driven particle crossing but do not allow direct measurement of true particle interpenetration or disentangle mixing from heating and spectral evolution.

Observational studies have reported mixed magnetosheath and magnetospheric particle populations within KH vortices and boundary layers, together with signatures of reconnection and VIR (e.g., \citealt{hasegawa2004,stawarz2016,nakamura2017a,settino2022,radhakrishnan2024}). These measurements provide compelling evidence that KH activity facilitates plasma entry. However, in situ observations are inherently local and do not directly quantify the degree of irreversible cross-field interpenetration over global magnetopause scales. Fully kinetic simulations can therefore help clarify the mechanisms and spatial localization of mixing within KH structures. These simulations provide the possibility to label particles according to their initial spatial origin, enabling a direct quantification of cross-boundary transport. Label-based mixing diagnostics have been introduced in previous PIC studies to investigate plasma entry during KH evolution. In particular, \citet{nakamura2014} performed three-dimensional PIC simulations and quantified plasma transport using a mixing fraction based on particle origin. In that configuration, secondary interchange-type instabilities developed and contributed to turbulent plasma transport across the boundary layer.

The present study examines a different kinetic regime, characterized by an in-plane magnetic field and a symmetric-shear configuration. We focus on how mixing develops during the nonlinear evolution of KH vortices under these conditions, quantifying the mixing spatial localization, species dependence, and temporal evolution, and relating it to kinetic reconnection diagnostics. Rather than estimating global transport rates, our goal is to determine where and how genuine particle interpenetration occurs and whether mixing remains localized at kinetic scales.

To this end, we combine particle labeling with a complementary density-based tracer and reconnection diagnostics to quantify the spatial structure of plasma interpenetration during the nonlinear stage of the KHI using fully kinetic Particle-in-Cell (PIC) simulations of an electron--ion plasma. This framework enables a direct comparison between particle-origin mixing metrics and reconnection activity, allowing us to assess whether and to what extent magnetic reconnection is associated with localized cross-field transport. In contrast to previous studies that primarily inferred plasma transport from large-scale topology changes or global mixing fractions, we focus on the spatial localization and species dependence of genuine particle interpenetration at kinetic scales. By combining particle-origin labeling, a complementary density-based tracer, and reconnection diagnostics, we provide a quantitative assessment of how and where magnetic reconnection activity coincides with enhanced cross-field mixing during the nonlinear evolution of the KHI.


\begin{figure*}[t]
\centering
\includegraphics[width=\textwidth]{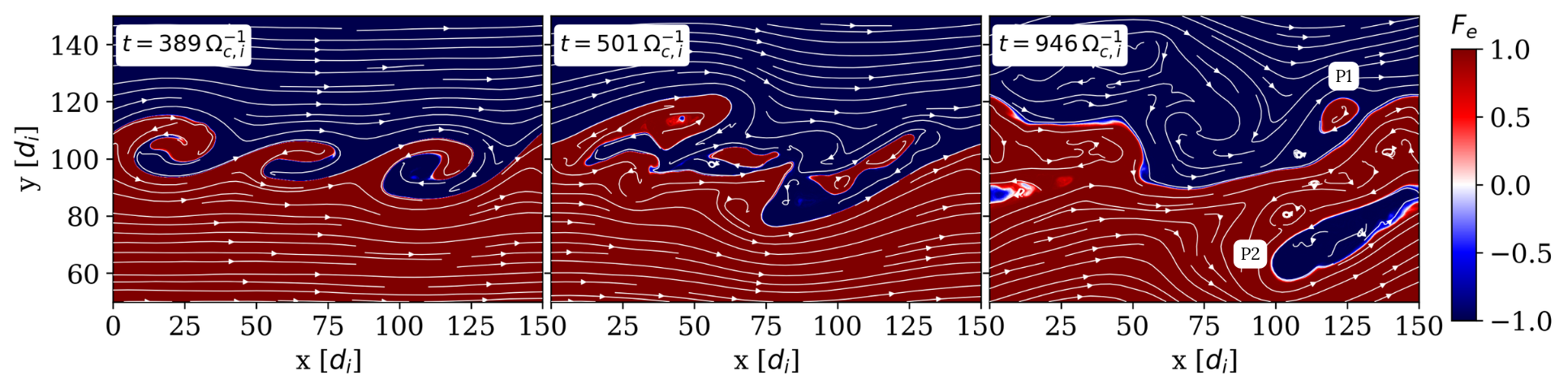}
\caption{\label{fig:tevol_mixing}From left to right: electron mixing fraction $F_e$ in the lower shear layer during the nonlinear phase. Shown are the rolled-up vortices at $t = 389\,\Omega_{c,i}^{-1}$, their merging at $t = 501\,\Omega_{c,i}^{-1}$, and the late nonlinear stage at $t = 946\,\Omega_{c,i}^{-1}$. White lines trace the in-plane magnetic field; in the right panel, two ejected plasma parcels (P1: red; P2:blue) are highlighted.}
\end{figure*}

\section{Methodology and Simulation Setup}
\label{sec:method}
We perform 2D (PIC) simulations using semi-implicit methods (the Implicit Moment Method and the Energy Conserving Semi-Implicit Method) implemented in the iPIC3D and ECsim codes (\citealt{markidis2010,lapenta2017a,bacchini2023,croonen2024}).  
Details of the numerical methods are given in \autoref{appendix:PICmethods}.

We model a collisionless magnetized plasma with uniform per-species density $n_0$ and an initial magnetic field composed of an in-plane component along $x$ and a guide field $B_{z,0} = 10 B_{x,0}$. All quantities are normalized using ion (mass $m_i$, charge $q_i$) reference scales: the ion gyro-frequency $\Omega_{c,i} = q_i B_\mathrm{tot,0}/(m_i c)$, where $B_\mathrm{tot,0} = \sqrt{B_{x,0}^2 + B_{z,0}^2}$ is the total magnetic field magnitude at initialization; the ion plasma frequency $\omega_{p,i} = \sqrt{4 \pi n_0 q_i^2 / m_i}$; the ion inertial length $d_i = c / \omega_{p,i}$; and the Alfvén speed $v_A = B_\mathrm{tot,0} / \sqrt{4 \pi n_0 m_i}$. At initialization, the ion cyclotron-to-plasma frequency ratio is $\Omega_{c,i}/\omega_{p,i} = 0.05$.

The initial equilibrium configuration is implemented following the approach of \cite{cerri2013}, ensuring a self-consistent shear-velocity profile within a finite-Larmor-radius (FLR) framework. 
Although this is not an exact kinetic equilibrium, it retains first-order FLR corrections, which significantly reduce the spurious fluctuations that arise from pressure-tensor responses at ion kinetic scales when using fluid-like MHD initial conditions, thereby yielding a more accurate representation of shear layers only a few ion Larmor-radii thick \citep{henri2013}.
Ions and electrons are initialized with drifting Maxwellian distributions 
with an initial drift velocity ${V}_\mathrm{drift} = 0.28 \,c$. Our boundary conditions are periodic in all directions.
We impose an initial velocity profile with a double shear 
\begin{equation}
    V_x(y) = V_{\text{drift}} \biggr[ \tanh \left( \frac{y - y_{sh,1}}{\delta} \right) - \tanh \left( \frac{y - y_{sh,2}}{\delta} \right) - 1 \biggr],
\end{equation}
where  $\delta=3d_i $ represents the initial shear-layer thickness, and the shears are located at $y_{sh,1}=L_y/4$ and $y_{sh,2}=3L_y/4$. The shear layers are chosen to study configurations where vorticity and guide magnetic field are parallel (at $y_{sh,1}$) or anti-parallel (at $y_{sh,2}$).
The computational domain is a 2D box of size $150d_i \times 400 d_i$ with $2304 \times 6144$ grid points, giving a spatial resolution $\Delta x = 0.0651\,d_i=0.5208\,d_e= 5.21\,\lambda_D$ with $\lambda_D = v_{th,e}/\omega_{p,e}$ the Debye length.
The time step is $\Delta t = 0.1 \omega_{p,i}^{-1}$. This choice resolves the electron gyration with approximately $17.6$ time steps per cyclotron rotation. The ion-to-electron mass ratio is $m_i/m_e = 64$, with 640 particles per cell. The temperature ratio is $T_e/T_i = 0.2$, and the plasma-$\beta = n_0 kT/(B^2_0/(8\pi))$ for ions and electrons is $\beta_i = 0.5$ and $\beta_e = 0.1$, respectively. 
The most unstable KH mode has a wavelength of $L_x/3$, leading to the formation of three vortices in each shear layer as the system enters the nonlinear phase. We identify four distinct stages in the evolution: a linear phase ($t=0$–$250\Omega_{c,i}^{-1}$) with exponential growth, an early nonlinear phase ($t=250$–$450\Omega_{c,i}^{-1}$), a merging phase ($t=450$–$650\Omega_{c,i}^{-1}$) dominated by vortex coalescence, and a late nonlinear phase ($t>650\Omega_{c,i}^{-1}$) characterized by increasingly turbulent behavior. In this study, we focus on the nonlinear evolution, including the early, merging, and late nonlinear phases. 

Plasma mixing is tracked by assigning each particle a label $\ell$ based on its initial $y$-position, with $\ell=0$ for particles initialized in $y<y_{sh,1}$ or $y>y_{sh,2}$, and $\ell=1$ for those initialized in $y_{sh,1}<y<y_{sh,2}$.
This labeling allows us to distinguish particles originating from different regions and to track their movement across shear layers, providing a new measure of plasma transport and mixing for both ions and electrons that isolates genuine interpenetration from coherent vortex advection and by systematically quantifying its spatial and temporal evolution at kinetic scales.

\section{Results: Plasma Mixing and Magnetic Reconnection}
In \autoref{fig:tevol_mixing}, we show snapshots representative of the evolution of the vortices at the lower shear during the nonlinear stage of the KHI. The quantity shown is the electron mixing fraction (\citealt{nakamura2014}),
\begin{equation}
F_e = \frac{(n_{e,\ell=1}-n_{e,\ell=0})}{n_e},
\end{equation}
where $n_{e,\ell=1}$ and $n_{e,\ell=0}$ are the electron number densities initially on either side of the shear. Regions with $F_e = -1$, $F_e = 1$, and $|F_e| < 1$ correspond to unmixed plasma on either side and mixed plasma, respectively. The ion mixing fraction and the upper shear layer for both species behave similarly and are not shown. 

The left panel of \autoref{fig:tevol_mixing} shows three rolled-up vortices at the early nonlinear stage. As they merge (middle panel), a dynamic \textit{mixing layer} forms, where plasma parcels can temporarily cross regions before returning to their origin (\autoref{fig:tevol_mixing}, middle and left panels). Magnetic tension limits permanent cross-field transport: although reconnection allows plasma parcels to detach from their original field lines, the strong in-plane field pulls them back. Consequently, magnetic tension counteracts plasma transport and forces macroscopic parcels to rebound toward their region of origin, limiting their ability to permanently migrate into the opposite region. In \autoref{appendix:inplanefield} we further elaborate on this conjecture, showing how the presence of an in-plane magnetic field influences the development of the KHI and prevents plasma from mixing efficiently.

While the mixing fraction $F_e$ provides the qualitative picture described above, it does not allow us to adequately quantify actual mixing because it does not capture the movement of small groups of particles potentially crossing between regions. To quantify interpenetration more accurately, we use a new tracer $\tilde{n}$ based on the labeling scheme described in \autoref{sec:method}. For each particle species, it is defined as the absolute change in particle density on either side of the shear:
\begin{equation}
\tilde{n}(\bb{x},t) = \sum_\ell \big| n_\ell(\bb{x},t) - n_\ell(\bb{x},t=0) \big|,
\end{equation}
summing over all labeled populations. This highlights regions where plasma from different initial regions mixes, providing a clearer visualization than the mixing fraction. In essence, this tracer identifies any cell of the domain in which particles with a certain label $\ell$ (belonging to a specific side of the shear at $t=0$) are present at time $t$. In regions where particles with a certain $\ell$ were not initially present, the tracer assumes a value $>0$.

\begin{figure*}[t]
\centering
\includegraphics[width=\textwidth]{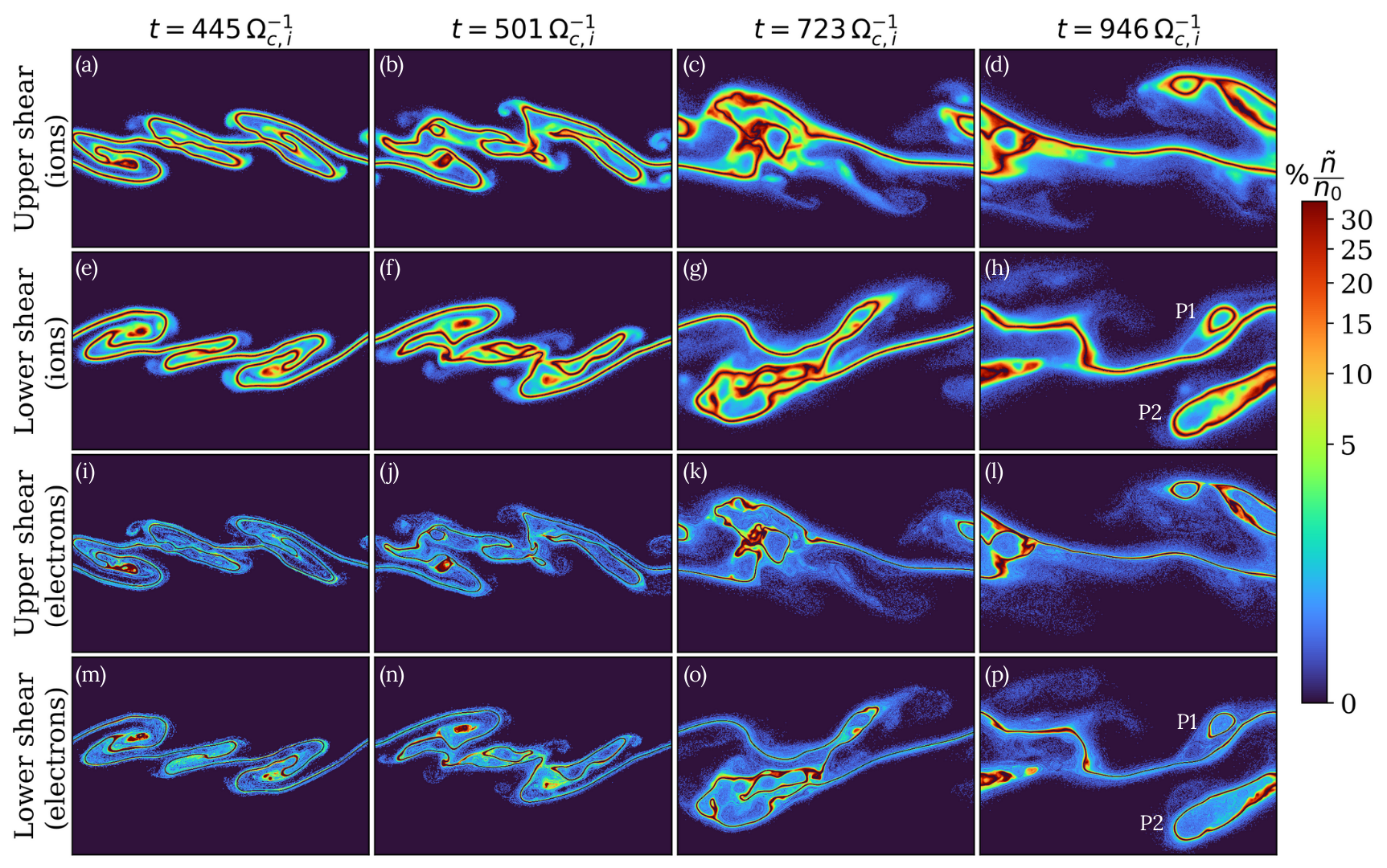}
\caption{\label{fig:tevol2D_dens_mixing} Morphological evolution of the lower ($y=y_{sh,1}$) and upper ($y=y_{sh,2}$) shear layers at $t=445,501,723,946 \, \Omega_{c,i}^{-1}$ during the nonlinear stage of the KHI. Each panel shows a region around each shear that is $100 \times 150 \, d_i^2$. The colorbar shows $\tilde{n}/n_0$ in percentage, which describes the mixing of the two plasma regions normalized to the initial value (see text).
The top two rows show the ion mixing (a--d upper shear, e--h lower shear), and the bottom two rows show the electron mixing (i--l upper shear, m--p lower shear). 
}
\end{figure*}

In \autoref{fig:tevol2D_dens_mixing}, we compare the spatial distribution of $\tilde{n}/n_0$ around both shear layers at different stages of the KHI for ions and electrons. In the top two rows, we observe that ions mix more efficiently both close to and farther away from the frontier. In the bottom two rows, we observe that electrons mix less efficiently, especially at early times. At later stages, there are localized regions around plasma parcels and the frontier where the electron and ion mixing level increases, reaching values up to $30\%$. These peaks are very localized, and in general, plasma mixing remains only mildly efficient, with most particles penetrating the opposite plasma region only marginally. For ions, higher mixing levels are observed, particularly within plasma parcels that have fully crossed from one region to the other. In the right panel of \autoref{fig:tevol_mixing}, two such parcels, P1 and P2, are highlighted and also shown in panels (h) and (p) of \autoref{fig:tevol2D_dens_mixing}, exhibiting higher internal mixing, particularly for ions in (h). This suggests that while plasma elements are advected coherently by the KH dynamics, intra-parcel diffusion promotes additional mixing. Other regions farther away from the frontier also show elevated mixing (see all panels in \autoref{fig:tevol2D_dens_mixing} at times $t=723, \; 946 \; \Omega_{c,i}^{-1}$).

\begin{figure}[t]
\centering
\includegraphics[width=\columnwidth]{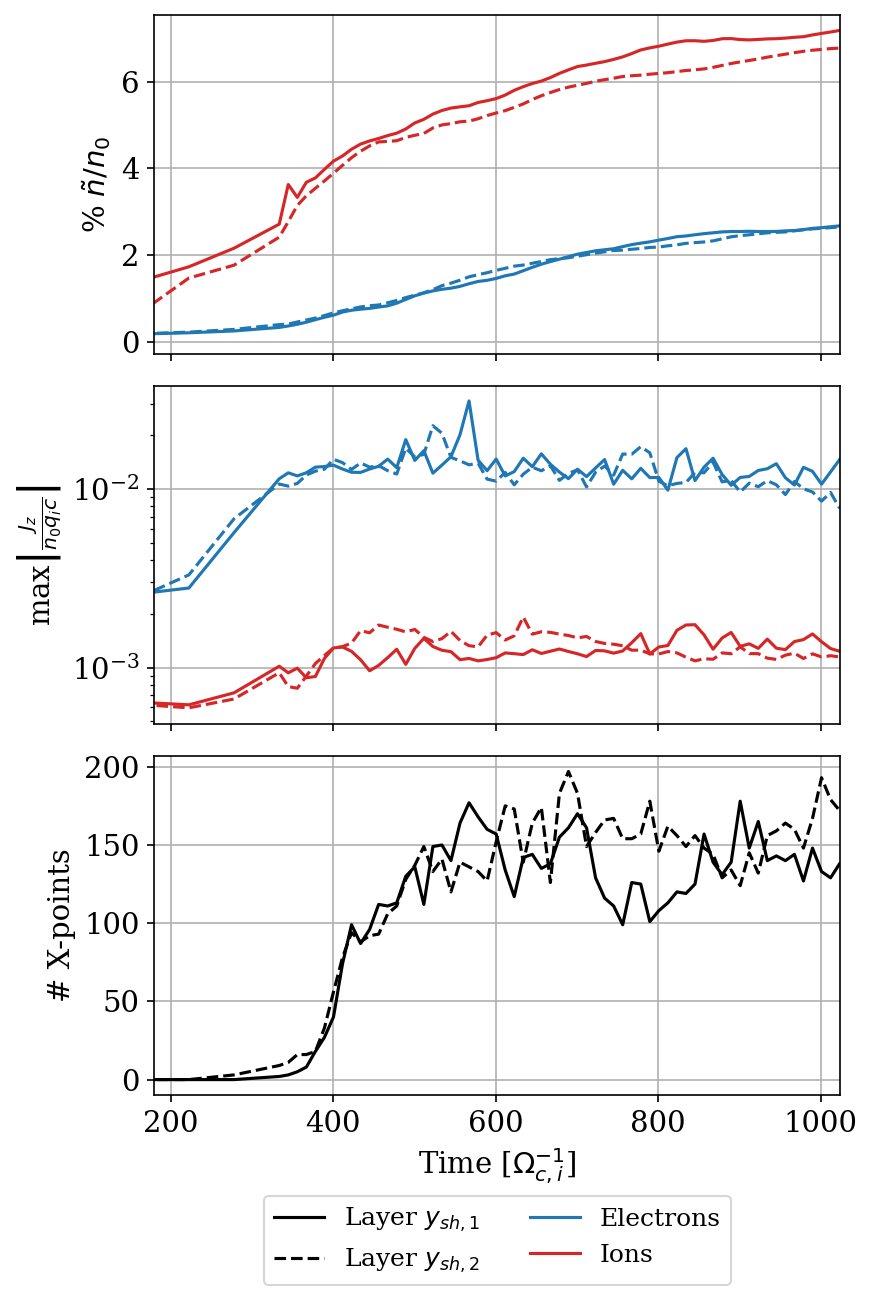}
\caption{\label{fig:evol_xpoints_jz_mixing} Temporal evolution of plasma mixing, current density, and X-points. Bottom and top shear layers ($y_{sh,1}$, solid; $y_{sh,2}$, dashed) are indicated in all panels. Top panel: percentage of mixed plasma $\tilde{n}/n_0$ for electrons (blue) and ions (red). Middle panel: maximum absolute out-of-plane current density $|J_z/(n_0 q_i c)|$ for electrons (blue) and ions (red). Bottom panel: number of X-points in each shear layer.}
\end{figure}

To quantify plasma mixing, we compute the volume-averaged $\tilde{n}/n_0$ within the $100 \times 150\; d_i^2$ regions highlighted in \autoref{fig:tevol2D_dens_mixing}. The top panel of \autoref{fig:evol_xpoints_jz_mixing} shows its temporal evolution in both shear layers during the nonlinear phase.
Although mixing can locally reach $30\%$, its strong localization keeps the volume-averaged fraction low. Ions cross the interface more efficiently than electrons, reaching averages of $\sim 7\%$ compared to $\sim 3\%$ for electrons. Around $t \simeq 300 \, \Omega_{c,i}^{-1}$, when fully rolled-up vortices form, mixing increases for both species, especially ions, then plateaus after $t \simeq 850 \, \Omega_{c,i}^{-1}$.
As \autoref{fig:tevol2D_dens_mixing} shows, low average mixing is expected, since mixing mostly occurs near the interface and within plasma parcels crossing it coherently. We also show here that mixing is similar for different layers, implying that anti--alignment between vorticity and magnetic field plays no evident role in this dynamics.

To understand the link between magnetic-field dynamics and mixing, we analyzed magnetic-reconnection activity by measuring the out-of-plane current density and X-point formation at the shear layers. Magnetic reconnection has previously been identified as a key process in the nonlinear evolution of KH vortices and suggested as a potential driver of plasma transport (e.g., \citealt{hasegawa2004,nakamura2011}). Here, we directly quantify its association with particle interpenetration using our mixing tracer.
The middle panel of \autoref{fig:evol_xpoints_jz_mixing} shows the temporal evolution of the maximum out-of-plane current density $J_z$ for each species and shear layer. In our 2D geometry, the reversal of the in-plane magnetic field is sustained by a $J_z$ sharply localized at X-points, making it an effective proxy for magnetic reconnection. $J_z$ evolves similarly in both shears, with electron currents generally larger in magnitude, as expected from the mass difference between species. For $t\gtrsim 200\ \Omega_{c,i}^{-1}$, the current density begins to increase in both species, reaching saturation around $t\simeq 400\ \Omega_{c,i}^{-1}$. This coincides with the onset of the early nonlinear KHI and the beginning of the vortex-merging stage (left and middle panels of \autoref{fig:tevol_mixing}), as well as the previously noted rise in the mixing percentage.
The bottom panel of \autoref{fig:evol_xpoints_jz_mixing} shows the number of X-points over time at each shear layer. X- and O-points are identified as nulls of the magnetic-flux gradient, $\nabla \psi = \mathbf{0}$, and distinguished by the sign of the Hessian determinant, following standard 2D reconnection diagnostics (\citealt{servidio2009,wan2013,olshevsky2016}). The X-point count increases from $t\gtrsim 300\ \Omega_{c,i}^{-1}$, corresponding to the early nonlinear stage of the KHI.
These diagnostics indicate that enhanced reconnection activity—higher $\max|J_z|$ and more X-points—coincides with increased plasma mixing ($\tilde{n}/n_0$). As reconnection breaks and reconnects field lines, particles from opposite sides of the shear are transported along newly connected lines, promoting cross-field mixing in frontier regions and detached plasma parcels. The spatial correspondence between X-point clusters and regions of elevated mixing, together with their synchronized temporal evolution, provides strong evidence that magnetic reconnection facilitates the localized mixing observed during the nonlinear development of the KHI.

\section{Conclusions}
We investigated the efficiency of plasma mixing driven by the nonlinear development of the magnetized KHI using high-resolution 2D fully kinetic simulations with finite-Larmor-radius effects (\citealt{cerri2013}) and diagnostics including particle labeling, a new mixing tracer, and reconnection proxies. Unlike previously introduced observational mixing parameters (\citealt{settino2022,settino2024}), which rely on energy-space separation of particle populations, our diagnostics are based on direct origin tracking in a fully kinetic framework. This enables a quantitative assessment of true particle interpenetration and its localization at electron scales, providing a complementary perspective to spacecraft-based mixing measures. Our results show that, in this setup, mixing between magnetosheath and magnetospheric plasmas is present but relatively inefficient, even after KH vortices form and the nonlinear stage develops. Mixing is highly localized, reaching $~30\%$ only in sparse locations within a narrow interface layer and isolated advected parcels, and is generally absent farther from the shear layers. On average, ions achieve somewhat higher mixing levels than electrons, whereas electron mixing remains weak and largely confined to localized structures. We can explain this behavior by considering that ions are significantly less magnetized than electrons, allowing them to partially cross the shear interface, whereas electrons remain tightly bound to field lines. With a reduced mass ratio ($m_i/m_e=64$), the actual electron magnetization in physical plasmas would be even higher, and their mixing correspondingly lower, meaning that electron mixing here likely represents an upper bound. Enhanced mixing occurs mainly within plasma parcels that detach and reattach across the interface, suggesting a significant contribution from magnetic reconnection and individual particle dynamics. Regions located between the interface and these parcels exhibit intermittent increase in mixing, coinciding with enhanced magnetic reconnection signatures, such as elevated out-of-plane current and higher X-point counts. We find that enhanced mixing preferentially occurs in regions where reconnection signatures are present, suggesting that magnetic reconnection facilitates localized cross-field particle interpenetration during the nonlinear development of the KHI.

While these results provide new insights into localized plasma transport during the nonlinear KHI, several limitations should be noted. The simulations are two-dimensional and use a reduced ion-to-electron mass ratio, which may overestimate electron mobility and restrict the range of fully three-dimensional reconnection dynamics. 
In addition, the present simulations adopt a symmetric configuration across the shear layer, whereas the Earth's magnetopause is typically asymmetric in density, temperature, and magnetic-field strength. Asymmetric configurations can trigger additional interchange-type instabilities and enhance turbulent transport \citep{nakamura2014,dargent2019}. The symmetric setup considered here isolates the role of velocity shear, magnetic geometry, and magnetic reconnection in regulating cross-field transport. Our results should therefore be interpreted as characterizing plasma mixing in a magnetized, symmetry-controlled regime, providing a reference case against which more realistic asymmetric configurations can be assessed. While strongly asymmetric conditions may enhance turbulent transport and increase the net amount of cross-boundary plasma exchange, the mechanisms identified here—localized mixing near reconnection sites and within advected plasma parcels—are expected to remain key ingredients of kinetic KHI evolution in asymmetric configurations. 
The finite horizontal extent of the simulation domain limits the ultimate growth of individual vortices due to periodic boundary conditions. In the reference configuration, vortex merging leads to structures approaching the box size in $x$, which prevents further large-scale horizontal expansion. However, we run additional simulations with larger domains confirming that increasing the number of vortices does not qualitatively alter the localized nature of mixing at kinetic scales. Although longer evolution times could increase the cumulative transport, the strong in-plane magnetic field constrains vertical broadening of the mixing layer, and interpenetration remains confined to narrow regions near reconnection sites and detached plasma parcels. These results should therefore be interpreted as characterizing the kinetic regulation of mixing within KH vortices rather than the long-term global evolution of the magnetopause boundary layer.

Despite these limitations, this study provides clear, fully kinetic evidence that plasma mixing driven by the nonlinear KHI can remain intrinsically localized and strongly influenced by in-plane magnetic-field components. By introducing a complementary mixing tracer and combining it with particle labeling and reconnection diagnostics, we identify the specific structures—detached plasma parcels and reconnection sites—where cross-field transport is enhanced. Our results demonstrate that enhanced mixing preferentially coincides with magnetic-reconnection signatures, establishing a quantitative connection between kinetic reconnection activity and localized plasma interpenetration in nonlinear KH dynamics.

Future work will extend these studies to fully three-dimensional setups, explore realistic mass ratios, and conduct a more detailed investigation of reconnection physics at the particle level, which could further clarify its role in facilitating cross-field transport.

\begin{acknowledgements}
S.F.\ is supported by the FWO PhD fellowship "Investigating Magnetospheric Plasma Dynamics with Large--scale Fully Kinetic Simulations" (grant no.\ 1126325N). 
F.B.\ acknowledges support from the FED-tWIN programme (profile Prf-2020-004, project ``ENERGY'') issued by BELSPO, and from the FWO Junior Research Project G020224N granted by the Research Foundation -- Flanders (FWO). 
The resources and services used in this work were provided by the VSC (Flemish Supercomputer Center), funded by the Research Foundation - Flanders (FWO) and the Flemish Government.
\end{acknowledgements}

\bibliographystyle{aa}
\bibliography{biblio}

\appendix  


\section{Semi-Implicit PIC Methods}
\label{appendix:PICmethods}

PIC codes typically use explicit time integration, which requires resolving the Debye length and the electron plasma frequency and is constrained by the Courant--Friedrichs--Lewy condition. Semi-implicit methods relax these constraints by linearizing or exactly treating the coupling between particles and fields, allowing simulations to retain small temporal and spatial scales, including electron-scale dynamics, while employing large domains. In this work, the first part of the simulation uses the Implicit Moment Method (IMM, \citealt{brackbill1982}), which approximates the particle--field coupling and provides enhanced stability compared to explicit methods. The subsequent stage employs the Energy Conserving Semi-Implicit Method (ECSIM, \citealt{lapenta2017a}). ECSIM replaces the nonlinear particle--field coupling with an exact relation, conserves energy exactly, retains the stability of fully implicit methods, and allows larger time steps and grid sizes while accurately capturing electron-scale dynamics. 

The results shown in this work correspond to the nonlinear stage of the Kelvin--Helmholtz instability. We start the simulation with the IMM until the KHI develops, and then switch to ECSIM at the early stages of the nonlinear phase, so that all results presented here are obtained using the ECSIM method.

\section{Effect of the In-plane Magnetic Field on Plasma Mixing}
\label{appendix:inplanefield}

\begin{figure}
\centering
\includegraphics[
width=\columnwidth
]{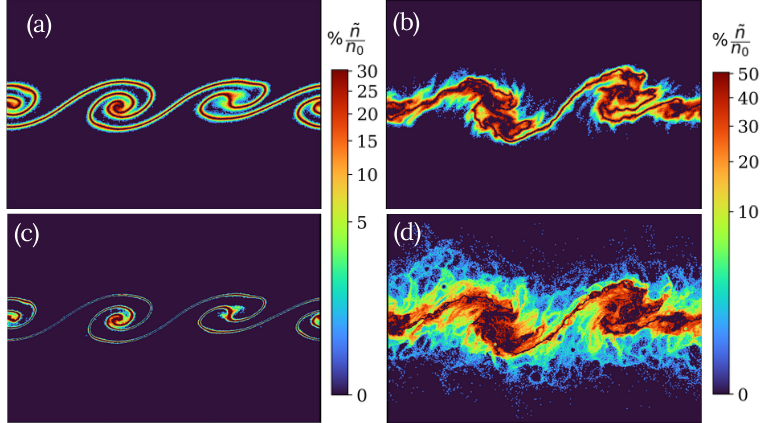}
\caption{\label{fig:MRvsMRNOT}
Panels (a) and (c) show the percentage of mixed plasma, $\tilde{n}/n_0$, for ions and electrons in the lower shear layer with a finite $B_x$ at $t = 334\,\Omega_{c,i}^{-1}$. Panels (b) and (d) show the corresponding ion and electron distributions in the lower shear layer without $B_x$ at the same time.
Each panel shows a region around the shear layer of size $100 \times 150 \, d_i^2$. A power-law normalization is used to enhance the visibility of low-amplitude structures.
}
\end{figure}
To further support the interpretation proposed in the main text, we compare in \autoref{fig:MRvsMRNOT} the nonlinear evolution of the lower shear layer in simulations performed with and without an in-plane magnetic field $B_x$. These simulations employ the same physical and numerical setup as the reference run discussed in \autoref{sec:method}, but with a spatial resolution reduced by a factor of four. Despite the reduced resolution, the simulations capture the large-scale nonlinear dynamics of the KHI and enable a clear assessment of the role of the in-plane magnetic field in regulating plasma mixing.

Panels (a) and (c) of \autoref{fig:MRvsMRNOT} show the spatial distribution of the percentage of mixed plasma, $\tilde{n}/n_0$, for ions and electrons, respectively, in the presence of a finite $B_x$. In this case, as KH vortices develop, plasma mixing remains weak and is largely confined to the narrow interface between the two plasma regions. Although vortical structures form and advect plasma across the shear layer, the in-plane magnetic field limits cross-field transport, resulting in well-defined vortices and localized mixing (up to $30\%$) primarily concentrated near the frontier, in line with the discussion presented in the main text. In contrast, panels (b) and (d) show the corresponding ion and electron distributions, always for the lower shear, this time in the absence of an in-plane magnetic field. In this case, the nonlinear evolution of the KHI is accompanied by significantly enhanced plasma mixing (up to $50\%$), with particles transported far from the initial interface and mixing regions extending well into both plasma domains. The absence of magnetic tension allows vortical motions to more efficiently redistribute plasma across the shear layer, leading to a broader and more spatially extended mixing region.

This comparison indicates that the limited and localized mixing observed in the main simulations is a robust physical effect associated with the presence of the in-plane magnetic field, rather than a consequence of numerical resolution. The in-plane field thus plays a key role in constraining cross-field transport and limiting the efficiency of plasma mixing in the nonlinear phase of the instability.

\end{document}